\documentclass[preprint,10pt,sort&compress]{elsarticle}
\usepackage{amssymb,amsfonts,amsmath}
\usepackage{graphicx}
\pdfoutput=1

\begin{document}
\renewcommand{\thefootnote}{\fnsymbol{footnote}}
\title{Probing Scalar Coupling Differences via Long-Lived Singlet States}

\author[hc]{Stephen J. DeVience}
 
\author[hs,hp,hcbs]{Ronald L. Walsworth\corref{cor1}}
\ead{rwalsworth@cfa.harvard.edu}

\author[hp,hms,mc]{Matthew S. Rosen}

\address[hc]{Department of Chemistry and Chemical Biology, Harvard University, 12 Oxford St., Cambridge, MA 02138, USA.}
\address[hs]{Harvard-Smithsonian Center for Astrophysics, 60 Garden St., Cambridge, MA 02138, USA.}
\address[hp]{Department of Physics, Harvard University, 17 Oxford St., Cambridge, MA 02138, USA.}
\address[hcbs]{Center for Brain Science, Harvard University, 52 Oxford St., Cambridge, MA 02138, USA.}
\address[hms]{Harvard Medical School, 25 Shattuck Street, Boston, MA 02115, USA.}
\address[mc]{A. A. Martinos Center for Biomedical Imaging, 149 Thirteenth St., Charlestown, MA 02129, USA.}
\cortext[cor1]{Corresponding author\\
Address:\\
Harvard-Smithsonian Center for Astrophysics,\\
MS 59, 60 Garden St., Cambridge, MA 02138}

\begin{abstract}
We probe small scalar coupling differences via the coherent interactions between two nuclear spin singlet states in organic molecules. We show that the spin-lock induced crossing (SLIC) technique enables the coherent transfer of singlet order between one spin pair and another. The transfer is mediated by the difference in {\em cis} and {\em trans} vicinal J couplings among the spins. By measuring the transfer rate, we calculate a J coupling difference of 8$\pm$2 mHz in phenylalanine-glycine-glycine and $2.57 \pm 0.04$ Hz in glutamate. We also characterize a coherence between two singlet states in glutamate, which may enable the creation of a long-lived quantum memory.
\end{abstract}

\maketitle

\section{Introduction}
Magnetic resonance experiments utilizing hyperpolarized media or exploring weak spin-spin interactions are often limited by the relatively short spin-lattice relaxation time, $T_1$, of the nuclear spins. Recently, the $T_1$ limit was greatly extended for a wide variety of molecules using nuclear spin singlet states, which can possess lifetimes $T_S > T_1$ \cite{Levitt1, Levitt5, Levitt6, Levitt7, Ghosh1, Bodenhausen4, Bodenhausen5, DeVience1}. The singlet state has subsequently been used to extend the lifetime of hyperpolarized spins, to measure slow diffusion and transport, and to isolate targeted NMR spectra in heterogeneous samples via a quantum filter \cite{Warren1, Bodenhausen2, Bodenhausen3, Bodenhausen6, Pileio3, Pileio5, DeVience3}. As a consequence of symmetry differences, the singlet state does not interact with its associated triplet states, and singlet state relaxation only occurs via weaker higher-order processes involving surrounding spins \cite{DeVience1, Pileio1, Pileio4, Levitt3, Levitt4}. However, the same symmetry differences also make the singlet state inaccessible via conventional RF pulses. An early solution for creating singlet states was to start with pairs of inequivalent spins, such as those of citric acid or alanine (Fig.\ 1a), and then create a singlet state using strong spin-locking. Subsequently, pulse sequences such as M2S and SLIC were developed to transfer polarization from the triplet states to the singlet state of two nearly equivalent spins by utilizing their small chemical shift difference, thereby making spin-locking unnecessary \cite{Levitt11,DeVience2,Ivanov2014}. Analogous techniques have been demonstrated for a particular 4-spin configuration of coupled spin pairs (Fig.\ 1b), which allows the singlet state of magnetically equivalent spins to be accessed, and for transferring polarization from parahydrogen to substrate molecules during signal amplification by reversible exchange (SABRE) polarization \cite{Theis1,Warren2,Claytor201481,Zhang1,Theis2,Pravdivtsev2}. In all cases, the pulse sequences can be used not only to prepare singlet states, but also to determine spectroscopic parameters, such as J coupling and chemical shift differences, that would otherwise be hidden in a conventional spectrum due to spin state dressing.

In the symmetric 4-spin systems previously studied, the singlet state cannot be prepared selectively in one spin pair or the other. Instead, spin order is simultaneously transferred from two triplet states into two singlet states. Although that configuration creates ideal singlet populations for hyperpolarized work, in certain situations it would be useful to create a larger variety of spin states. For example, if one wishes to store quantum information in the form of arbitrary populations and coherences of long-lived singlet states, then each individual singlet state must be made accessible. To explore this possibility, we consider here a 4-spin configuration in which the two distinct singlet states in two nearly equivalent spin pairs of an organic molecule can be prepared and manipulated selectively  (Fig.\ 1c). We show that coherent interactions occur between the two singlet states mediated by the difference in {\em cis} and {\em trans} vicinal J couplings. This interaction allows small J coupling differences to be measured, even when $\Delta J \ll 1/T_1$. Moreover, the singlet-singlet interaction can be controlled and used to create a coherent superposition between two singlet states.

We demonstrate the technique in proton pairs of the glutamate molecule, in which we selectively prepare one singlet state and then perform both Rabi and Ramsey measurements in a subspace defined by the two singlet states. In the Rabi measurement, singlet state polarization is coherently transferred back and forth between the two spin pairs. The rate of transfer provides a measure of singlet-singlet interaction strength. In the Ramsey measurement, a coherent superposition of singlet states is created and allowed to evolve in the singlet-singlet subspace before being projected onto one spin pair or the other. The frequency of the associated oscillations provides a measurement of the energy difference between singlet states. We also implement a Rabi measurement in the phenylalanine-glycine-glycine (phe-gly-gly) molecule to demonstrate the technique's utility for measuring weak coupling differences. In phe-gly-gly, the scalar coupling between spin pairs is small compared with the spin-spin relaxation rate, i.e., $J \ll 1/T_2 \approx 1/T_1$, so that the coupling is difficult to measure from spectra or coherence transfer experiments. However, because the nuclear spin singlet state has a lifetime $T_S \gg T_1$, coherent interactions between singlet states are detectable, and we are able to measure a scalar coupling difference on the order of 10 mHz among the spins.

\begin{figure}
\centering
\includegraphics[width=3.54in]{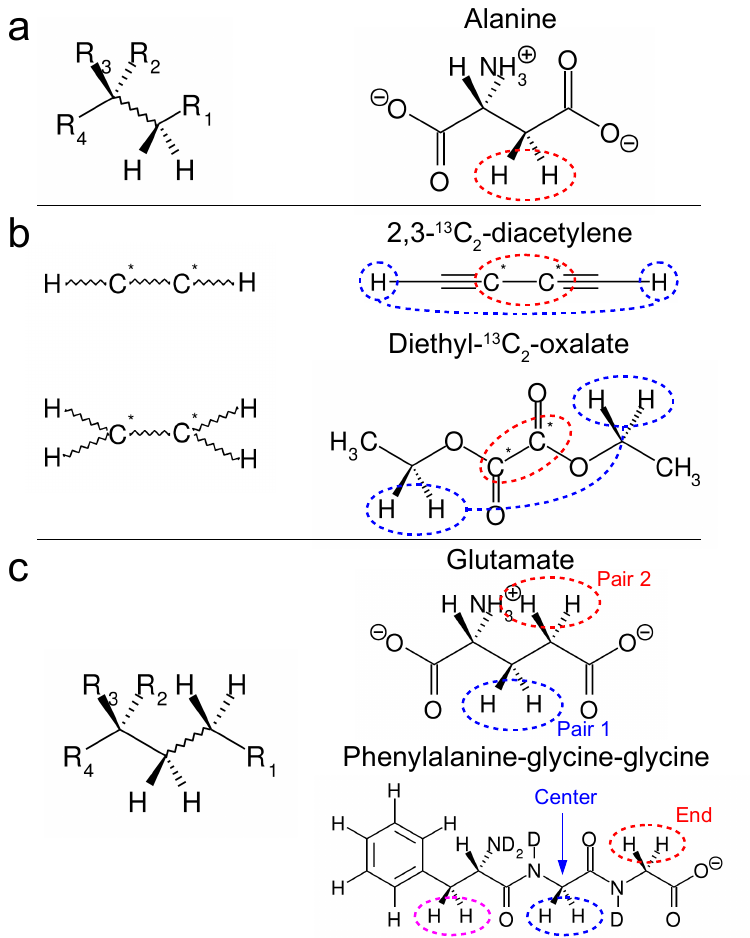}  
\caption{Example families of molecular structures used for singlet state experiments. Squiggly lines indicate that a variable number of intervening bonds is possible. (a) A pair of inequivalent or nearly equivalent geminal protons. Examples: alanine, citric acid, glycerol formal. (b) One pair of identical spins sandwiched between one or more pairs of identical spins of a different nucleus (here * indicates locations enriched with $^{13}$C). Examples: diacetylene, diethyloxalate. (c) Two or more pairs of nearly equivalent protons. Examples: glutamate, phenylalanine-glycine-glycine. Other geometries not shown include vicinal proton pairs, neighboring $^{15}$N nuclei, and neighboring phosphate groups.}
\label{fig:structures}
\end{figure}

\section{Theory}

For a single spin in a static magnetic field $B_0$, the spin eigenstates can be described in the Zeeman basis as aligned or anti-aligned with $B_0$, i.e., $\vert \uparrow \rangle$ or $\vert \downarrow \rangle$. However, in a system with multiple spins, the wavefunctions can only be described exactly in the Zeeman basis in the case of non-interacting spins. Any spin-spin interactions cause the Zeeman states to mix, thereby creating dressed states that represent the true eigenstates of the system. In the case $\vert J \vert < \vert \Delta \nu \vert$, i.e., weak scalar coupling compared with any resonance frequency difference between spins, the dressed eigenstates can be calculated using perturbation theory, and they remain substantially similar to the Zeeman eigenstates. However, in the case $\vert J \vert \gg \vert \Delta \nu \vert$, perturbation theory fails and the dressed states must instead be calculated by diagonalizing the full Hamiltonian. The resulting eigenstates are very different from the Zeeman states and can possess unique properties, such as an extended lifetime in the case of the singlet state.

One of the simplest multiple spin systems consists of two spin-1/2 nuclei interacting via scalar coupling. If the coupling strength is zero, the spin states can be represented by the product states $\vert \uparrow \uparrow \rangle, \vert \uparrow \downarrow \rangle, \vert \downarrow \uparrow \rangle$, and $\vert \downarrow \downarrow \rangle$. When the scalar coupling between spins is strong compared with their resonance frequency difference ($\vert J \vert \gg \vert \Delta \nu \vert$), or if strong spin-locking is applied ($\vert \nu_n \vert > 5 \vert \Delta \nu \vert$, where $\nu_n$ is the spin-lock nutation frequency), then diagonalization reveals that the spin pair is instead described by singlet and triplet eigenstates. For the case of strong scalar coupling and magnetically equivalent spins, one finds a spin-0 singlet state $ \vert S_0 \rangle = (\vert \uparrow \downarrow \rangle - \vert \downarrow \uparrow \rangle)/\sqrt{2}$ and three spin-1 triplet states $ \vert T_{-} \rangle = \vert \uparrow \uparrow \rangle$, $ \vert T_0 \rangle = (\vert \uparrow \downarrow \rangle + \vert \downarrow \uparrow \rangle)/\sqrt{2}$, and $ \vert T_{+} \rangle = \vert \downarrow \downarrow \rangle$.

If the two spins are nearly but not exactly equivalent, then the small chemical shift difference, $\Delta \nu$, couples singlet and triplet states. As described elsewhere, this interaction can be used to prepare the singlet state by spin-locking at a nutation frequency such that triplet and singlet energy levels cross (spin-lock induced crossing or SLIC) \cite{DeVience2,Theis1,Feng1,Pravdivtsev1}. The effect comes about because spin-locking rearranges the triplet states and perturbs their energy levels, leading to the spin-locked eigenstates:
\begin{align}
&\vert \phi _{+}\rangle = \frac{1}{2}(\vert \uparrow \downarrow \rangle + \vert \downarrow \uparrow \rangle+\vert \uparrow \uparrow \rangle + \vert \downarrow \downarrow \rangle) = \frac{1}{\sqrt{2}}\vert T_0 \rangle + \frac{1}{2}(\vert T_{-} \rangle+\vert T_{+} \rangle)\nonumber\\
&\vert \phi _{0}\rangle = \frac{1}{\sqrt{2}}(\vert \uparrow \uparrow \rangle - \vert \downarrow \downarrow \rangle)=\frac{1}{\sqrt{2}}(\vert T_{-} \rangle - \vert T_{+} \rangle)\nonumber\\
&\vert \phi _{S}\rangle = \frac{1}{\sqrt{2}}(\vert \uparrow \downarrow \rangle - \vert \downarrow \uparrow \rangle) = \vert S_0 \rangle\nonumber\\
&\vert \phi _{-}\rangle = \frac{1}{2}(\vert \uparrow \downarrow \rangle + \vert \downarrow \uparrow \rangle-\vert \uparrow \uparrow \rangle-\vert \downarrow \downarrow \rangle) = \frac{1}{\sqrt{2}}\vert T_0 \rangle - \frac{1}{2}(\vert T_{-} \rangle+\vert T_{+} \rangle).
\end{align} 

In a system composed of two pairs of spin-1/2 nuclei, the pairs can be labeled $i$ = 1 and 2, with each pair's spins labeled $j$ = $a$ and $b$. For the molecules studied, the interpair spin-spin coupling is weak compared with the intrapair coupling, so at first order the system can be approximated by 16 new product states formed from the singlet and triplet states of each spin pair. The relationships among these states form the basis for previous studies in which polarization was transfered from triplet product states $\vert T T \rangle$ to the singlet product state $\vert S_0 S_0 \rangle$ \cite{Theis1,Claytor201481,Warren2}. In the present study, we consider a different situation, in which we selectively transfer polarization to the singlet state of one spin pair while the other spin pair remains in the triplet state. We then only need to consider product states containing the triplet states of one spin pair and the singlet state of the other, represented by $\vert T S_0\rangle $ and $\vert S_0 T\rangle$. Here, the overall triplet component is some combination of the individual triplet states and can be described by
\begin{equation}
\vert T \rangle_{i} = \alpha_i \vert \phi_{+} \rangle_i + \beta_i \vert \phi_{0} \rangle_i + \gamma_i \vert \phi_{-} \rangle_i,
\end{equation}
where $\alpha_i$, $\beta_i$, and $\gamma_i$ are real amplitudes, and to preserve normalization we require
\begin{equation}
\alpha_i^2 + \beta_i^2  + \gamma_i^2 = 1.
\end{equation}

Couplings between the states $\vert T S_0\rangle $ and $\vert S_0 T\rangle$ arise because the spins of pair 1 are also weakly coupled to the spins of pair 2, i.e., there are couplings $J_{1a2a}$, $J_{1a2b}$, $J_{1b2b}$, $J_{1b2a}$. These couplings produce an interaction between the states with strength
\begin{equation}
C = \frac{J_{1a2a}+J_{1b2b}-J_{1a2b}-J_{1b2a}}{4}\left(\alpha_1 \alpha_2 + \beta_1 \beta_2 + \gamma_1 \gamma_2 \right).
\end{equation}
Note that the interaction is mediated by the difference in $cis$ and $trans$ J-couplings within the molecule, where $cis$ refers to two nuclei on the same side of the molecular backbone and $trans$ refers to two nuclei on the opposite side. Although intramolecular configurations change rapidly in the molecules studied, there is a different average coupling $J_{iai'a}$ versus $J_{iai'b}$ no matter which particular spins are labeled $a$ and $b$ \cite{Karplus1, Karplus2}. The antisymmetric form of the interaction is necessary to couple the two antisymmetric eigenstates. The sum of $cis$ and $trans$ couplings leads to interactions among the symmetric product states.

The energy of each state is a function of the spin-lock nutation frequency for each spin pair, $\nu_{n,1}$ and $\nu_{n,2}$, and the intrapair J couplings:
\begin{align}
&E_1 = \frac{J_{1a1b}}{4}-\frac{3 J_{2a2b}}{4}+(\alpha_1^2 -\gamma_1^2) \nu_{n,1} \nonumber \\
&E_2 = \frac{J_{2a2b}}{4}-\frac{3 J_{1a1b}}{4}+(\alpha_2^2 -\gamma_2^2 )\nu_{n,2} .
\end{align}
A Hamiltonian for $\vert T S_0\rangle $ and $\vert S_0 T\rangle$ can then be written as
\begin{equation}
\mathcal{H} = h \left[ \begin{array}{cc}
E_1 & C  \\[0.5em]
C & E_2 
\end{array}\right].
\end{equation}
The interaction terms have no effect unless $\vert \Delta E \vert = \vert E_1 - E_2 \vert < \vert C \vert $, i.e., the interactions are ineffective at driving transitions as long as the coupling terms are smaller than the energy difference between the eigenstates. For $\nu_{n,1} = \nu_{n,2} = 0$, this energy difference is simply $\vert \Delta E \vert = \vert J_{1a1b} - J_{2a2b} \vert$. However, a subset of states can be brought on resonance by spin-locking, which modifies the energy of those states containing $\alpha$ or $\gamma$ terms. For example, if spin-locking is applied resonant with spin pair 1, then the difference in effective spin-lock nutation frequencies can be approximated by
\begin{equation}
\Delta \nu_n \approx \sqrt{\nu_{n,1}^2 + \Delta \nu_{12}^2}-\nu_{n,1},
\end{equation}
where $\Delta \nu_{12}$ is the average resonance frequency difference between spin pair 1 and spin pair 2. The RF spin-locking transmitter frequency and power can be chosen such that $\vert \Delta \nu_n \vert = \vert \nu_{n,1} - \nu_{n,2} \vert = \vert \Delta E \vert$. In that case, the two states are brought on resonance and coherent polarization transfer occurs, which results in the singlet state population of each spin pair exhibiting an oscillation with period
\begin{equation}
\tau = \frac{2}{J_{1a2a}+J_{1b2b}-J_{1a2b}-J_{1b2a}} = \frac{1}{J_{cis}-J_{trans}},
\end{equation}
where $J_{cis}$ and $J_{trans}$ are the average $cis$ and $trans$ scalar couplings. The transfer can also be thought of as a unitary transformation in a subspace consisting of two singlet states. Controlling the length of the transfer allows one to produce the equivalent of a $\pi$ pulse:
\begin{equation}
\vert \phi_{+} S_0 \rangle \rightarrow \vert S_0 \phi_{+} \rangle,
\end{equation}
and a $\pi/2$ pulse:
\begin{equation}
\vert \phi_{+} S_0 \rangle \rightarrow \frac{1}{\sqrt{2}}\left(\vert \phi_{+} S_0 \rangle+\vert S_0 \phi_{+}\rangle\right).
\end{equation}

The transfer of singlet state polarization or the creation of a singlet-singlet coherence can then be detected by measuring the singlet component of each spin pair, as discussed in the ``Experiment'' section. Although the triplet state is not measured, triplet-triplet interactions and relaxation affect the details of the transfer process. For instance, if the triplet states interconvert more slowly than $1/\tau$, then the interacting singlet states will be entangled with either $\vert \phi_{+} \rangle$ or $\vert \phi_{-} \rangle$, depending on the spin-locking phase (a 180$^\circ$ phase shift changes the sign of $\nu_n$). One could subsequently perform the operations $\vert \phi_{+} S_0 \rangle \rightarrow \vert S_0 \phi_{+} \rangle$ followed by $\vert \phi_{-} S_0 \rangle \rightarrow \vert S_0 \phi_{-} \rangle $ in two independent steps. On the other hand, if the triplet states interconvert more quickly than $1/\tau$, then the triplet states will always remain in equilibrium with one another and the singlet transfer will appear independent of the spin-locking phase. In that case the triplet states play a purely ancillary role by bringing the singlet levels into resonance, and it is possible to view the system as a subspace defined by two singlet states.

If the two spin pairs are sufficiently off-resonance, then when one spin pair is spin-locked the other experiences the interaction term $\nu_1 \hat{I}_{1z} + \nu_2 \hat{I}_{2z}$, which connects states $\vert \phi_{+} \rangle$ and $\vert \phi_{-} \rangle$. Consequently, either spin-locking phase effectively drives the transition
\begin{equation}
\vert \phi_{+} S_0 \rangle + \vert \phi_{-} S_0 \rangle \rightarrow \vert S_0 \phi_{+} \rangle + \vert S_0 \phi_{-} \rangle.
\end{equation}
Whether entanglement is preserved can be determined experimentally, as described below. Alternatively, the system can be analyzed by further dressing with the interaction $\nu_1 \hat{I}_{1z} + \nu_2 \hat{I}_{2z}$. Simulations show that this term produces energy shifts making it possible to drive singlet transfer even when the effective spin-locking is equal for both spin pairs.

\section{Experiment}

We used a 200 MHz Bruker NMR spectrometer to perform experiments on two samples: (1) an 80 mM solution of monosodium glutamate dissolved in pH 7.0 phosphate buffer and (2) a 20 mM solution of phenylalanine-glycine-glycine (phe-gly-gly) dissolved in D$_2$O. For both samples, nitrogen gas was bubbled through the solution for 5 minutes to displace dissolved oxygen. A reference spectrum was acquired for each sample using a $90^{\circ}$ pulse immediately followed by a FID acquisition. Measurements of $T_1$ were performed with an inversion recovery sequence.

Next, singlet state preparation sequences were calibrated for each molecule. For glutamate, SLIC was used for singlet state preparation and readout of both spin pairs, and calibration of the transfer time and spin-locking nutation frequency were performed as described in \cite{DeVience2}. For phe-gly-gly, the singlet state of the nearly equivalent spin pair was accessed via SLIC, while the singlet state of the second spin pair was accessed via a three-pulse sequence and preserved by spin-locking \cite{Bodenhausen4, Levitt1, DeVience1}. For all cases, we find that the target singlet state can be produced selectively in one of the two proton pairs, because the scalar coupling and resonance frequency are sufficiently different for each spin pair.

\begin{figure*}
\centering
\includegraphics[width=5.51in]{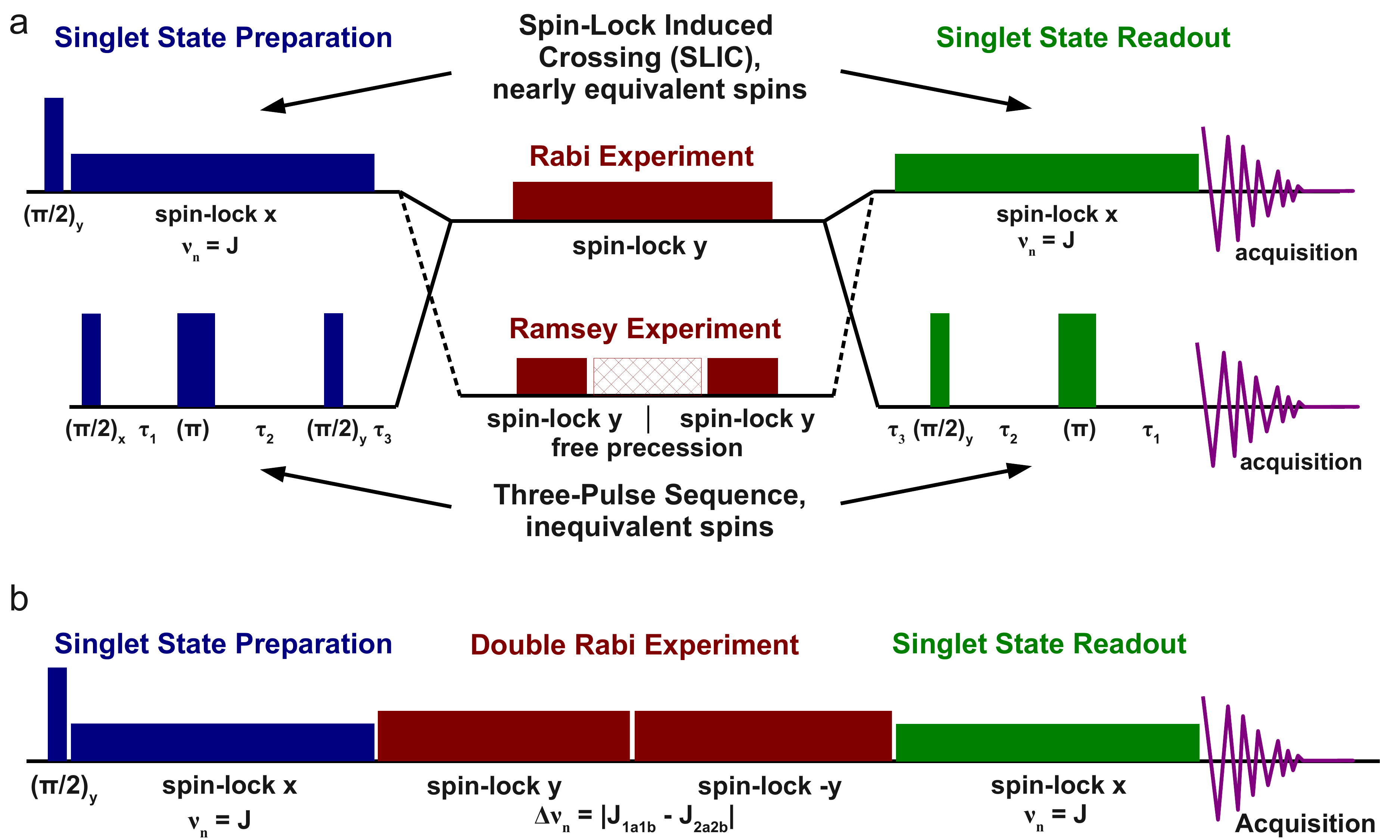}  
\caption{(a) Pulse sequences for singlet state transfer measurements. A singlet state is selectively created on one spin pair with either a spin-lock induced crossing (upper) or a three-pulse sequence (lower). A Rabi measurement can be performed to detect the singlet transfer rate to the second spin pair, or a Ramsey measurement can be performed to study a singlet-singlet coherence. Finally, the singlet state is read from one spin pair or the other before a FID signal is acquired. (b) Double Rabi sequence for determining whether entanglement is lost between the singlet state of one spin pair and the individual triplet states of the second spin pair.}
\label{fig:pulse_sequence}
\end{figure*}

We then implemented the pulse sequence shown graphically in Fig.\ 2a to measure the singlet-singlet interactions. Each sequence consists of three stages: singlet state preparation, singlet transfer operations, and singlet state readout. Preparation and readout were performed with SLIC or the three-pulse sequence, depending on the targeted spin pair's J coupling and chemical shift parameters. In a Rabi experiment, the target singlet state was prepared on one spin pair, CW spin-locking was applied to drive singlet transfer to the second spin pair, and the singlet state from either the first or second spin pair was converted back to transverse magnetization for readout using either a SLIC or three-pulse sequence. Measurements were performed in which either the spin-locking duration, $\tau_{\rm{SL}}$, or spin-locking nutation frequency, $\nu_{n}$, was varied while the other was kept constant. In a Ramsey experiment, the singlet state was prepared in one spin pair, and then spin-locking with a fixed nutation frequency and duration (calibrated with the Rabi experiment) was applied to create a coherent superposition between the two singlet states. This was followed by free precession of the singlet states for time $\tau_{\rm{Ramsey}}$ and then a subsequent spin-locking application to project the superposition onto the second spin pair. During the free precession period, spin-locking was applied at a power sufficient to prevent degeneracy of the triplet states but weak enough to avoid inducing singlet transfer. Finally, a SLIC or three-pulse readout sequence was selectively applied to one of the proton pairs to convert its singlet state population to transverse magnetization for readout. In both Rabi and Ramsey sequences, the transverse magnetization was measured by acquiring the FID signal. Phase cycling was applied to remove residual triplet magnetization \cite{Levitt1,Pileio2}.

To determine whether the singlet state of one spin pair maintained entanglement with a triplet state of the other pair, or whether the triplet states mixed and entanglement was lost, we performed the experimental protocol shown in Fig.\ 2b. During the Rabi sequence, spin-locking was first applied with phase $y$ for duration $\tau_{\rm{SL}}$, and then applied a second time with phase $-y$ for duration $\tau_{\rm{SL}}$. If entanglement was maintained, then when $\tau_{\rm{SL}} = 1/(2 \vert J_{\rm{cis}} - J_{\rm{trans}} \vert)$, maximal singlet state transfer would occur for each triplet case $\vert \phi_{+} \rangle$ and $\vert \phi_{-} \rangle$, and the singlet state transfer to the second spin pair would be maximized. On the other hand, if entanglement was lost, then a full period of singlet state transfer would occur for time $\tau_{\rm{SL}} = 1/(2\vert J_{\rm{cis}} - J_{\rm{trans}} \vert)$, and the singlet state would be transferred to the second spin pair and back again by the end of the sequence.

\section{Results}

\subsection{Glutamate}

\begin{figure*}
\centering
\includegraphics[width=5.5in]{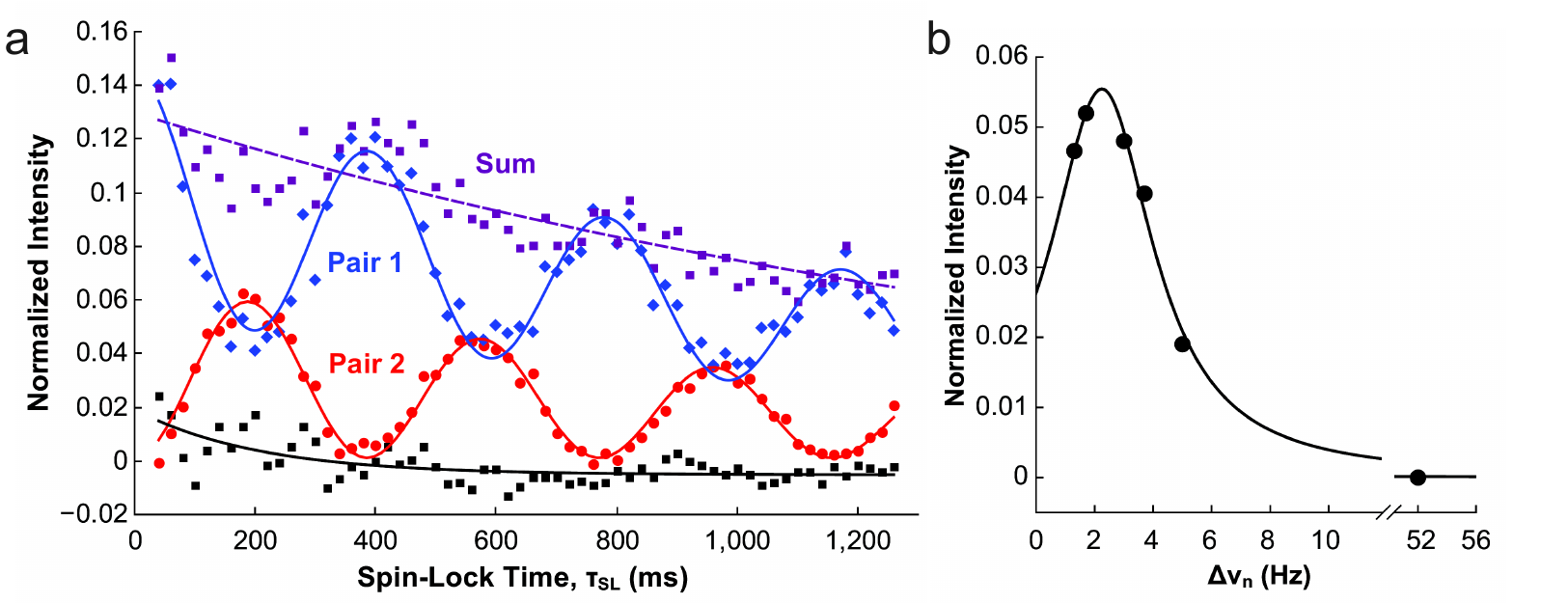}  
\caption[Measurement of coherent singlet state transfer in glutamate]{Measurement of coherent singlet state transfer in glutamate. (a) Singlet order is selectively prepared predominantly in spin pair 1 and is followed with spin-locking at a 500 Hz nutation rate during the evolution time. The singlet state is then read out from either spin pair 1 (blue points) or spin pair 2 (red points). Oscillations in the singlet state population of each spin pair indicate $f = \vert J_{\rm{cis}} - J_{\rm{trans}} \vert =  2.57 \pm 0.04$ Hz. If spin-locking is not applied, singlet transfer does not occur (black points). The small amount of residual magnetization decays with a time constant $T=205 \pm 7$ ms, indicating that it arises from short-lived triplet states. The sum of singlet state measurements (purple points) decays exponentially with time constant $T_{\rm{Rabi}} = 1.6 \pm 0.1$ s. All intensities are integrated signals normalized to a conventional spectrum. Spin pair 1 results were fit with a function $I(\tau_{\rm{SL}}) = A [\cos^2 (\pi f \tau_{\rm{SL}})+c]\exp(-\tau_{\rm{SL}}/T_{\rm{Rabi}})$ while spin pair 2 results were fit with a function $I(\tau_{\rm{SL}}) = A [\sin^2 (\pi f \tau_{\rm{SL}})+c]\exp(-\tau_{\rm{SL}}/T_{\rm{Rabi}})$. (b) The amplitude of singlet transfer, extracted from the parameter $A$ of the fit, is plotted for singlet transfer measurements performed with a range of spin-locking nutation frequencies. A Lorentzian fit gives a peak value of $2.25 \pm 0.08$ Hz for the resonance condition with a FWHM of $4.3 \pm 0.4$ Hz.}
\label{fig:glu_data}
\end{figure*}

The proton NMR spectrum of glutamate acquired at 200 MHz is shown in Fig.\ 3a. Each spin pair exhibits a multiplet pattern due to coupling with the second spin pair as well as coupling with another lone proton. Figure \ref{fig:glu_data}a shows results for a singlet transfer Rabi experiment in glutamate. The singlet state was prepared on spin pair 1 (chemical shift $\delta = 2.04$ ppm) using the SLIC sequence with $\nu_n = 15.5$ Hz and a spin-lock duration of 157 ms. Black points show the normalized integrated signal of spin pair 2 ($\delta =2.3$ ppm) when no spin-lock was applied ($\nu_{n} = 0$ Hz) during the transfer stage after reading out its singlet state using the SLIC sequence with $\nu_n = 17$ Hz and a spin-lock duration of 145 ms. While some residual magnetization was present, it decayed within $\sim$ 200 ms, indicating that it arose from short-lived triplet states and that no singlet state was created on spin pair 2. Moreover, we set an upper limit of $0.01$ for the possible amount of singlet transferred. Red points show the same measurement when a spin-lock with $\nu_{n} = 500$ Hz ($\Delta\nu_{n} = 2.7$ Hz) was applied during the transfer stage with the transmitter set to $\delta = 2.04$ ppm. The periodic oscillations indicate a coherent transfer of singlet state population between spin pairs 1 and 2. Blue points are results for the same experiment ($\nu_{n} = 500$ Hz, $\Delta\nu_{n} = 2.7$ Hz, transmitter at $\delta = 2.04$ ppm) when the singlet state was instead read out from spin pair 1. Notice that the oscillation has the same period but a $180^{\circ}$ phase shift, which indicates that singlet state from spin pair 1 was lost in proportion to that gained by spin pair 2. Approximately 1/3 of the singlet state from spin pair 1 was not transferred, as it corresponded to the population of spin pair 2 in the $\vert \phi_0 \rangle$ state. The period of the oscillation indicates a transfer frequency of $2.57 \pm 0.04$ Hz, which represents the average difference between $J_{\rm{cis}}$ and $J_{\rm{trans}}$. The total singlet state (purple points) exhibits a weak oscillation, indicating that there was also a small amount of coherent transfer into a state other than the singlet state of spin pair 2. During the Rabi experiment, the total singlet state population relaxed with a time constant $T_{\rm{Rabi}} = 1.6 \pm 0.1$ s, which is approximately 60\% longer than $T_1$ (measured $T_1$ values are $0.92\pm0.02$ s and $1.11\pm0.02$ s for spin pairs 1 and 2, respectively). The Rabi experiment was then repeated for a range of $\nu_{n}$ values during the singlet transfer stage and the amplitudes of the oscillations were measured (Fig.\ \ref{fig:glu_data}b). A best-fit Lorentzian indicates a resonance condition $\Delta \nu_n = \Delta E = 2.25 \pm 0.08$ Hz, where $\Delta \nu_n$ was calculated from $\nu_{n}$ via Eq.\ 7.

\begin{figure}
\centering
\includegraphics[width=3.54in]{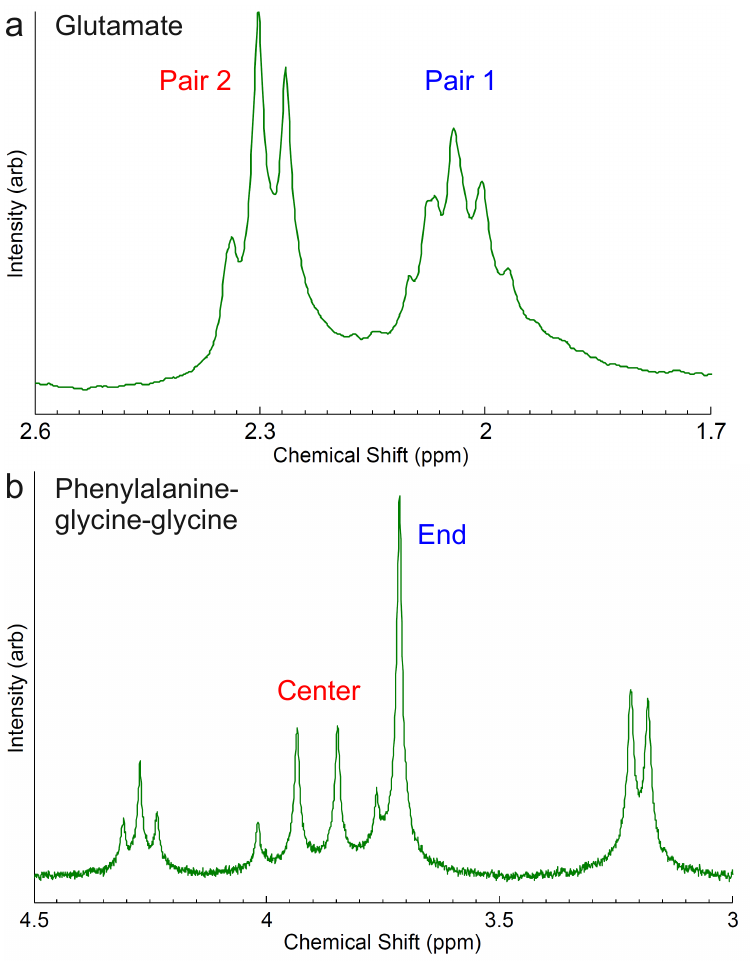}  
\caption{Example 200 MHz NMR spectra of (a) glutamate and (b) phenylalanine-glycine-glycine. Relevant peaks are labeled with their corresponding nuclei shown in Fig.\ 1. }
\label{fig:spectra}
\end{figure}

To perform a Ramsey experiment, we first prepared the singlet state in spin pair 1 and then performed a $\pi/2$ rotation in the singlet-singlet subspace by spin-locking with nutation frequency $\nu_{n} = 500$ Hz on-resonance with spin pair 1 ($\delta = 2.04$ ppm) for 100 ms. This created a coherent superposition state of the form $\left(\vert T S_0 \rangle+\vert S_0 T\rangle\right)/\sqrt{2}$. We next applied spin-locking for time $\tau_{\rm{Ramsey}}$ with nutation frequency $\nu_{n} = 47$ Hz ($\Delta \nu_n = 23$ Hz) to drive mixing among the triplet states without inducing singlet transfer. (We found that if spin-locking was not applied during the free precession period, we could not detect Ramsey oscillations.) We then performed a second $\pi/2$ rotation using identical spin-locking as for the first rotation, and we read out the singlet state from either spin pair 1 or 2 using SLIC. 

\begin{figure}
\centering
\includegraphics[width=5.5in]{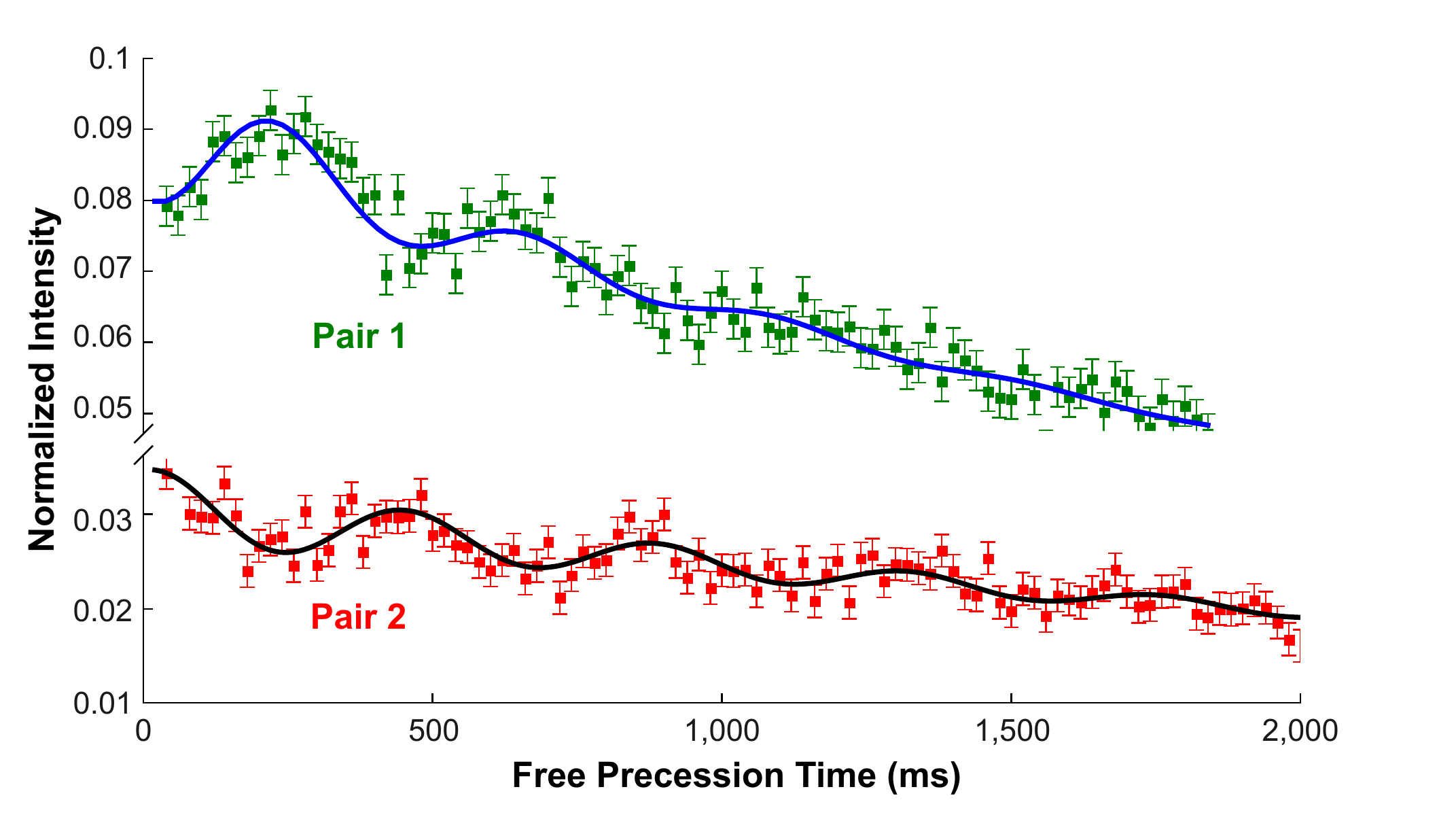}  
\caption[Measurements of Ramsey oscillations in the singlet-singlet subspace of glutamate]{Measurement of Ramsey oscillations in the singlet-singlet subspace of glutamate. Spin-locking is applied at nutation frequency $\nu_n = 47$ Hz with the transmitter centered on the pair 1 resonance frequency. Green and red points are measurements from spin pair 1 and 2, respectively. Black curves are sinusoidal fits with the function $I(\tau_{\rm{Ramsey}})=A [\cos(2 \pi f \tau_{\rm{Ramsey}}-\phi) \exp(-\tau_{\rm{Ramsey}}/T_{2S}^{*})+c] \exp(-\tau_{\rm{Ramsey}}/T_{S})$. Blue curves are sinusoidal fits with the function $I(\tau_{\rm{Ramsey}})=A [-\cos(2 \pi f \tau_{\rm{Ramsey}}-\phi) \exp(-\tau_{\rm{Ramsey}}/T_{2S}^{*})+c] \exp(-\tau_{\rm{Ramsey}}/T_{S})$. Intensity represents integrated signal normalized to a conventional 90$^{\circ}$-FID spectrum.}
\label{fig:glu_ramsey}
\end{figure}

Figure \ref{fig:glu_ramsey} shows results of the Ramsey experiment when the transmitter was centered on resonance with spin pair 1 during the free precession time. When the singlet state was read out from spin pair 2 (red points), the signal exhibited oscillations at $f = 2.33 \pm 0.03$ Hz, as calculated by fitting to the model function
\begin{equation}
I(\tau) = A [\cos(2 \pi f \tau_{\rm{Ramsey}}-\phi) \exp(-\tau_{\rm{Ramsey}}/T_{2S}^{*})+c] \exp(-\tau_{\rm{Ramsey}}/T_{S}).
\end{equation}
The oscillations represent the singlet-singlet coherent superposition precessing in the Bloch sphere defined by the singlet-triplet product states. We obtain a satisfactory fit ($\chi^2/$ degrees of freedom $\sim1$) when a small phase correction of $\phi = 20^{\circ}$ is applied. This shift may represent phase shifts between preparation and readout spin-locks or an initially lower precession frequency for initial time points due to an imbalance in the triplet populations (i.e., $\alpha \neq \gamma$). The signals exhibit two forms of decoherence: dephasing with a characteristic time $T_{2S}^{*}$ and depopulation with the singlet lifetime $T_{S}$. For readout from spin pair 2, we find values $T_{S}=4.4\pm 0.2$ s and $T_{2S}^{*} = 1.3 \pm 0.4$ s. When the singlet state was read out from spin pair 1 (green points), the signal exhibited corresponding oscillations $180^{\circ}$ out of phase with those from spin pair 2, with values $f = 2.30 \pm 0.08$ Hz, $T_{S}=2.91\pm 0.05$ s, and $T_{2S}^{*} = 0.55 \pm 0.13$ s. The frequency of both measurements roughly matches the singlet transfer resonance condition (Fig.\ 4b) and represents the energy difference between the two sets of singlet-triplet product states defining the Bloch sphere. The singlet state lifetimes were 2 to 4 times longer than $T_1$. The singlet state lifetime of spin pair 2 was longer than that of spin pair 1 because spin pair 2 is physically farther away from the downfield proton that drives singlet relaxation. For both cases, the measured dephasing time was $\sim 25\%$ of the singlet lifetime.

\begin{figure}
\centering
\includegraphics[width=5.5in]{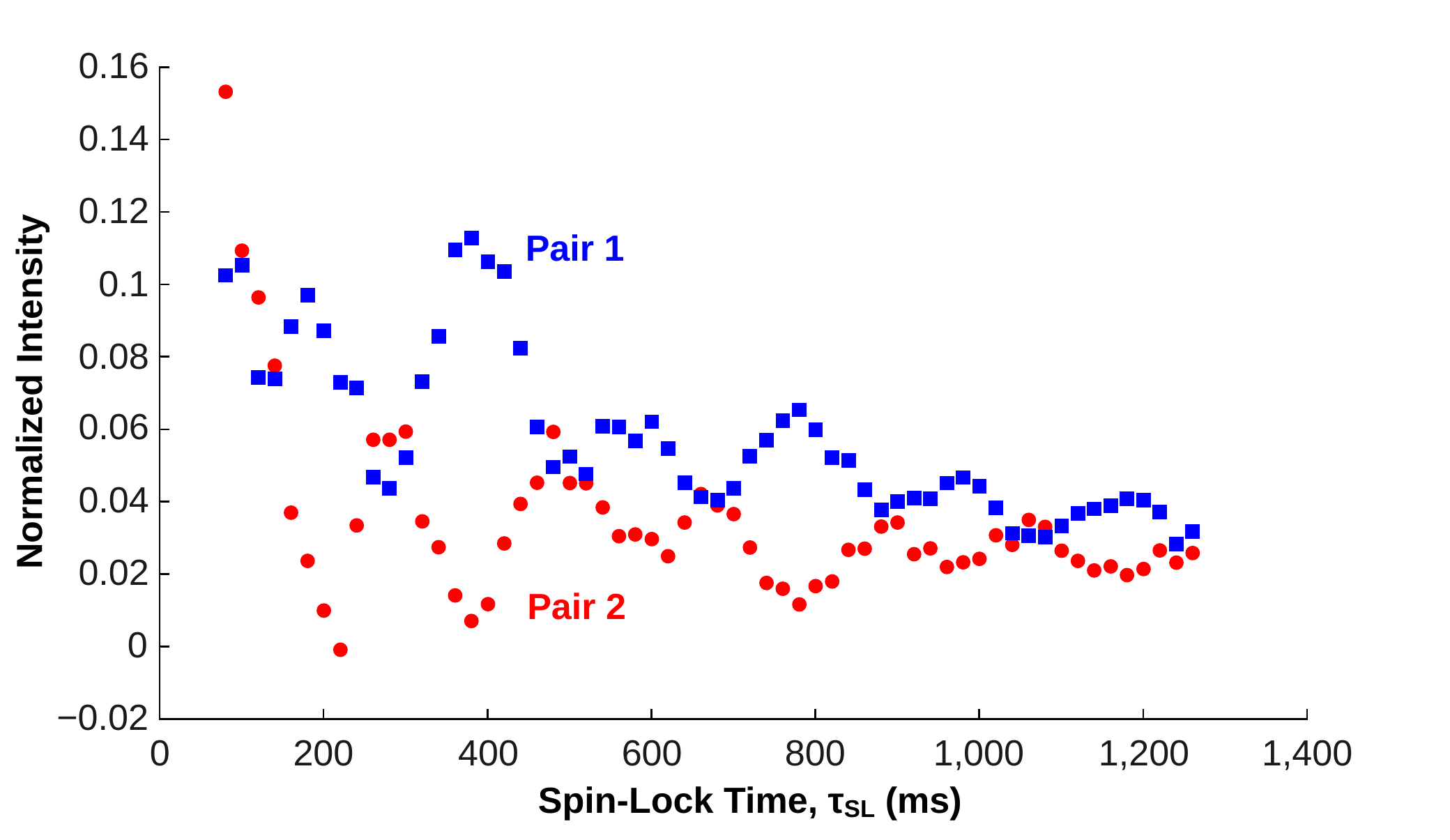}  
\caption[Measurement of singlet state transfer using the double Rabi experiment]{Measurement of singlet transfer using the double Rabi experiment. The spin-lock during the singlet transfer stage is applied for time $\tau_{\rm{SL}}$ twice, once with phase $y$ and once with phase -$y$. A full period of oscillation occurs when $2\tau_{\rm{SL}} = 1/\vert J_{\rm{cis}} - J_{\rm{trans}}\vert$, indicating that the transfer occurs simultaneously for both triplet components $\vert \phi_{+} \rangle$ and $\vert \phi_{-} \rangle$ rather than in separate stages.}
\label{fig:double_rabi}
\end{figure}

Finally, to confirm that the singlet state of spin pair 1 does not remain entangled with the triplet states of spin pair 2 after singlet state transfer, we performed the double Rabi experiment shown in Fig.\ 2b. A singlet state was prepared in spin pair 1, 500 Hz spin-locking was applied to spin pair 1 first with phase y and then with phase -y, and the singlet state was read out from either spin pair 1 or 2. The results (Fig.\ \ref{fig:double_rabi}) show oscillations with a period $T = 1/(2\vert J_{\rm{cis}} - J_{\rm{trans}} \vert)$, as expected if entanglement is lost. As a consequence, only one of the two spin-lock phases needs to be applied to transfer the full amount of accessible singlet state from one spin pair to the other. Moreover, as long as triplet states mix quickly enough, the triplet state component can be ignored and the Ramsey experiment can be viewed as taking place solely within a singlet-singlet subspace. The triplet states only act to make the singlet-singlet interactions accessible by bringing the states into resonance.

\subsection{Phenylalanine-glycine-glycine}

The spectrum of phenylalanine-glycine-glycine acquired at 200 MHz is shown if Fig.\ 3b. For measurements of singlet state transfer, we employed the pulse sequences of Fig.\ 2a, but we used the three-pulse sequence to prepare and read out the singlet state from the proton pair at $\delta = 3.89$ ppm (center of molecule, $\tau_1=7$ ms, $\tau_2=20.5$ ms, $\tau_3=9.25$ ms), while still using a SLIC sequence to access the proton pair at $\delta = 3.71$ ppm (end of molecule, $\nu_n = 17.5$ Hz, duration 300 ms). As in the glutamate Rabi experiment, a singlet state was first prepared in one of the two proton pairs, followed by a period of spin-locking for $\tau_{\rm{SL}}$ with an applied nutation rate $\nu_{n}$. The singlet state was then converted back to transverse magnetization in either the original proton pair or the adjacent pair before acquisition of the FID signal. 

\begin{figure*}
\centering
\includegraphics[width=7in]{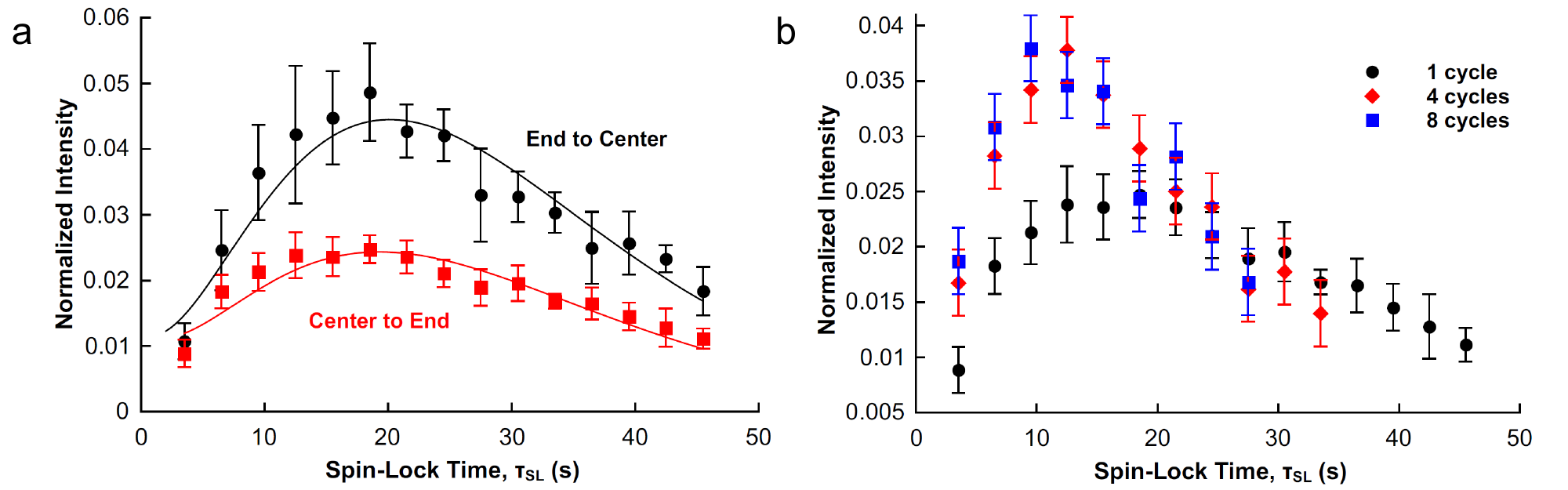}  
\caption[Measurement of singlet state transfer between proton spin pairs located at the center and end of the phenylalanine-glycine-glycine molecule]{Measurement of singlet state transfer in the phenylalanine-glycine-glycine molecule. (a) The singlet state is created in one spin pair and read out from the adjacent spin pair. Best-fit curves of Eq.\ 13 indicate a transfer rate $f = \vert J_{cis}-J_{trans} \vert = 8\pm2$ mHz. (b) The singlet state is created in the center spin pair and transferred to the end spin pair multiple times before readout in order to increase the total amount of singlet state transferred.}
\label{fig:phe-gly-gly-result}
\end{figure*}

Figure \ref{fig:phe-gly-gly-result}a presents measurements of singlet state transfer in phe-gly-gly between the center spin pair at $\delta = 3.89$ ppm and the end spin pair at $\delta = 3.71$ ppm when $\nu_{n} = 280$ Hz ($\Delta \nu_n = 2.3$ Hz).  Results are shown for two cases: (1) the singlet state was created in the end pair and read out from the center pair (black points), and (2) the singlet state was created in the center pair and read out from the end pair (red points). In both cases, the transmitter was set to $\delta = 3.89$ ppm for spin-locking to ensure a good lifetime for the spin-locked singlet. Notice that unlike the results for glutamate, no oscillation was observed, only a buildup of singlet state followed by a long decline. From a fit with the function
\begin{equation}
I(\tau_{\rm{SL}}) = A [\sin^2 (\pi f \tau_{\rm{SL}})+c]\exp(-\tau_{\rm{SL}}/T_{\rm{Rabi}}),
\end{equation}
we calculate a value $f = \vert J_{\rm{cis}}-J_{\rm{trans}} \vert = 8\pm2$ mHz and a decay time of $T_{\rm{Rabi}} = 11\pm3$ seconds, which is slightly less than the central spin pair's singlet lifetime $T_S = 14.5$ s. 

The transfer was less effective when transferring the singlet state from the center pair to the end pair, possibly because the singlet state can also be transferred in the other direction to the third spin pair of the chain. We attempted to increase the total amount of singlet state transferred using a ``pumping'' scheme to transfer the singlet state multiple times. We first created singlet state polarization on the center spin pair using the three-pulse sequence and spin-locked at $\nu_n = 280$ Hz to transfer the singlet state to the end pair. We then removed spin-locking for time $5 T_1 = 3.1$ s, created singlet state polarization in the center spin pair again using the three-pulse sequence, and repeated the spin-locking to induce a second singlet state transfer. This process was repeated multiple times before finally reading out the singlet state from the end pair using a SLIC sequence. After four of these pumping cycles, we were able to achieve a significant increase in the singlet state population of the end pair versus a single transfer sequence (Fig.\ \ref{fig:phe-gly-gly-result}b). Further increasing the number of cycles to eight had little effect, as the 25 s lifetime of the end spin pair's singlet state likely limited the total amount of singlet state buildup achievable.

\section{Discussion}

The above experimental results show that interactions between two singlet states within a molecule can produce coherent transfer of singlet order between one spin pair and another. The resonance condition needed to effectuate this transfer is controllable, allowing unitary operations to be performed in a subspace spanned by two singlet states. Because singlet state relaxation is slow compared to spin-lattice relaxation, this represents a step toward a decoherence-free subspace (DFS) in which quantum information can be stored beyond conventional relaxation times. A DFS consisting of four spins has been proposed before \cite{Lidar1, Antonio1, Bacon1}, using the two spin-zero states $\vert S_0 S_0 \rangle$ and $(\vert T_{+} T_{-} \rangle-\vert T_{0} T_{0} \rangle+\vert T_{-}  T_{+} \rangle)/\sqrt{3}$. Our system differs in that we utilize the individual singlet states of each spin pair. Decoherence within the DFS then occurs through depopulation of the singlet states as well as dephasing due to fluctuations in the singlet state energy levels. Since the singlet state energy levels are defined by the geminal J coupling between spins, they are sensitive to scalar relaxation of the first kind, which is generally extremely slow compared with other relaxation mechanisms \cite{Abragam1}. Our measurements of glutamate indicate a depopulation time greater than $T_1$ but a dephasing time similar to $T_1$, indicating that the geminal J coupling might be changing on a time scale similar to the free precession time. Dephasing can also be caused by fluctuations in the triplet state components of each spin pair if they do not interchange quickly enough relative to the free precession time.

We also demonstrated the measurement of weak J coupling differences on the order of 10 mHz. Such measurements could be useful for the determination of molecular structures, especially for proteins and macromolecules, as the difference in $cis$ and $trans$ J couplings is a function of molecular geometry \cite{Karplus1, Karplus2}. Many current multidimensional NMR methodologies for structure determination, such as COSY and TOCSY \cite{Aue1}, depend on coherence transfer among states that relax on the order of $1/T_2$ or $1/T_{1\rho}$ ($\approx 1/T_1$ for small molecules in solution). As proton spin lifetimes rarely exceed a few seconds, this sets a lower limit of the order 100 mHz for coupling resolution, which limits the ability to detect long-range coupling differences. Singlet state transfer may provide a way to detect weaker coupling differences that provide information about structure at distances five or more bond lengths apart, as well as allow for measurements within peptide chains such as phenylalanine-glycine-glycine, which contain only distantly-spaced protons.

\section{Acknowledgements}

This work was supported by the National Science Foundation and the Smithsonian Astrophysical Observatory.

\newpage
\bibliographystyle{elsarticle-num-names}
\bibliography{singlet_bib}
%

\end{document}